\documentstyle[12pt]{article}

\begin{document}

\def\be{\begin{equation}}
\def\ee{\end{equation}}
\def\bea{\begin{eqnarray}}
\def\eea{\end{eqnarray}}
\def\bb{\begin{array}{l}}
\def\ba{\begin{array}}
\def\ea{\end{array}}
\def\disp#1{\displaystyle{#1}}

\def\su#1#2{{\vbox{\hbox{$\scriptstyle #1$}
\vskip 3pt \hbox{$\scriptstyle #2$}}}}

\newtheorem{theorem}{Theorem}
\newtheorem{definition}[theorem]{Definition}
\newtheorem{proposition}[theorem]{Proposition}
\newtheorem{lemma}[theorem]{Lemma}
\newtheorem{conjecture}[theorem]{Conjecture}
\def\proof{\smallskip \par \noindent{\em Proof:\ }}
\def\remark{\smallskip \par \noindent{\em Remark:\ }}
\def\qed{{\ {\vrule height 3mm width 2mm depth 0 mm} \hfill}}
\def\lf{\nonumber \\ {} \nonumber \\ {}}
\def\llf{\\ {} \nonumber \\ {}}

\newcommand{\sect}[1]{
\setcounter{equation}{0}\setcounter{theorem}{0}\section{#1}}
\renewcommand{\theequation}{\thesection.\arabic{equation}}
\renewcommand{\thetheorem}{\thesection.\arabic{theorem}}

\def\Ga{\Gamma}
\def\la{\lambda}
\def\La{\Lambda}
\def\eps{\epsilon}
\def\si{\sigma}
\def\ot{\otimes}
\def\d{\partial}
\def\del{\nabla}
\def\im{\hbox{\rm im }}
\def\tr{\hbox{\rm tr }}
\def\vol#1{{| #1 |}}
\def\half{{1 \over 2}}

\def\and{, \;}

\def\lte{\prec}
\def\gte{\succ}

\def\TT{{\bf T}}

\def\IX{\hbox{\raise.2ex
\hbox{{$\scriptscriptstyle\mid$}}\kern-0.28em\hbox{{$\times$}}}}

\def\IR{\hbox{\rm{{I}\kern-0.2em{R}}}}
\def\IT{\hbox{\rm{{T}\kern-0.5em{T}}}}
\def\IZ{\hbox{\rm{{Z}\kern-0.4em{Z}}}}
\def\IN{\hbox{\rm{{I}\kern-0.2em{N}}}}
\def\one{\hbox{\rm{{1}\kern-0.34em{1}}}}
\def\stroke{\vrule height7pt width0.4pt depth-0pt}

\def\IC{\hbox{\rm {C} \kern-0.8em {\stroke} \kern0.2em{}}}

\def\TN{{\hbox{$\IT^N$}}}
\def\Ttwo{{\hbox{$\IT^2$}}}
\def\ZN{{\hbox{$\IZ^N$}}}

\def\vect{{vect(\TN)}}
\def\map{{map(\TN, \oj)}}
\def\fcn{{fun(\TN)}}

\def\tvect{{\widetilde{vect}(\TN; e)}}
\def\tvectb{{\widetilde{vect}(\TN; e, \beta)}}
\def\tvectbtwo{{\widetilde{vect}(\Ttwo; e, \beta)}}
\def\tfcn{{\widetilde{fun}(\TN; e)}}
\def\tfcnb{{\widetilde{fun}(\TN; e, \beta)}}

\def\ffd#1#2{f_{FD}(#1, #2)}

\def\oj{{\bf g}}
\def\hh{{\bf h}}
\def\toj{{\widetilde{\oj}}}

\def\emx{{e^{im \cdot x}}}
\def\enx{{e^{in \cdot x}}}
\def\ennx{{e^{-in \cdot x}}}

\def\ind#1#2{{{#1}_1 \cdots {#1}_{#2}}}
\def\indx#1#2#3{{{#1}_1 \cdots #3 \cdots {#1}_{#2}}}

\def\bra#1{{\langle #1 |}}
\def\ket#1{{| #1 \rangle}}
\def\sing{\ket{sing}}

\def\Fock{{\cal F}}
\def\psic{{\psi^\dagger}}
\def\limE{{\lim_{E \to \infty}}}

\def\tto{\longrightarrow}


\title{$\ZN$-graded Lie algebras: Fock representations and
reducibility conditions}

\author{T. A. Larsson\thanks{
Supported by the Swedish Natural Science Council (NFR)} \\
Dept. of Theoretical Physics \\
Royal Institute of Technology \\
100 44 Stockholm, Sweden
}
\date{December 1992}

\maketitle

\begin{abstract}
\noindent Manifestly consistent Fock representations of non-central
(but ``core-central'')
extensions of the $\ZN$-graded algebras of functions and vector fields
on the $N$-dimensional torus $\TN$ are constructed by a kind of
renormalization procedure.
These modules are of lowest-energy type, but the energy is not a
linear function of the momentum.
Modulo a technical assumption, reducibility conditions are proved
for the extension of $\vect$, analogous to the discrete
series of Virasoro representations.
\end{abstract}


\medskip
hep-th/9212055

\footnotetext{Email address: tl@theophys.kth.se}

\vfill \eject

\sect{Introduction}

A well-understood representation theory for algebras
of gauge transformations and vector fields exists only if the
dimension $N$ of the base manifold is at most one. To
obtain lowest-energy modules, the ``quantum version'' of an
algebra must be considered. If $N=1$ the quantum version
is a central extension, i.e.
the affine Kac-Moody and Virasoro algebras, respectively
\cite{GO,Kac,FF}.
If $N=0$, the current algebra is finite-dimensional, and it
has no extension at all.

However, little is known about the higher-dimensional case
\cite{Rud} -- \cite{Mic}. There are two problems. A na\"\i ve
attempt to construct Fock modules using normal ordering results
in an infinite central extension, which clearly signals an inconsistency
\cite{FR,Lar2}. Second, it is not clear what the quantum version is
in higher dimensions. It can not be a central extension, because
no interesting one exists \cite{RSS,Lar1}. Therefore, we
expect the quantum version to be something more complicated, and the
obvious suggestion is that it is a non-central extension. These have
appeared in physics in connection with anomalies \cite{Fad}, and
some representations have been constructed \cite{Mic}.

The problem with non-central extensions is that they come in large
numbers, but it is not easy to construct interesting modules
with a given extension.
Therefore, we suggest the opposite approach: construct interesting
modules and determine the extension afterwards.
In quantum physics relevant modules are characterized by the
energy being bounded from below.
We start with the Fock construction, but modify it where it goes wrong,
which leads to a modification of the algebra itself.
This can be regarded as a kind of renormalization.
Indeed, the normal ordering prescription can be described in the
same way:
the original vacuum energy is infinite, but when it is assigned
a finite value by hand, a central extension appears.
Normal ordering is part of our prescription, but we must also make
another redefinition to avoid other divergencies. The resulting
module is manifestly consistent and the energy is bounded from below.
The price is that we have a representation of a different
algebra, but we will argue that it has the correct ``classical limit''.
To describe this algebra in terms of generators and brackets
is very complicated and can probably not be done in closed form.
Nevertheless, its representations in Fock space are easily described.

In this paper we are concerned with the algebras $\fcn$ and $\vect$
of functions and vector fields on the $N$-dimensional torus $\TN$,
but the same methods can be used also for
current, Poisson and Moyal algebras on $\TN$.
These algebras will collectively be referred to as
{\em torus algebras}.
The restriction to the torus is of technical nature;
since every function can be decomposed into a Fourier series, these
algebras possess a natural $\ZN$-grading. However, since all
manifolds of the same dimensions are locally diffeomorphic, any local
result has wider applicability. Moreover, the formula
\be
\exp(i(m_1 x^1 + \ldots + m_N x^N)) = (t^1)^{m_1} \ldots (t^N)^{m_N},
\ee
establishes an isomorphism between plane waves and monomials, so
all results on $\TN$ hold for Laurent polynomials as well.

Lowest-weight modules of course exist for $\IN^N$-graded algebras,
such as the algebra of polynomials in $N$ variables, with the grading
given by the total degree of monomials.
However, this is not very interesting, because it is an algebra
with a lowest root, and also because no involution can be naturally
defined. These problems can be remedied by considering
Laurent polynomials, but in that case the homogeneous subspaces are
infinite-dimensional and we are back to our original problem.
Rudakov studied representations of $\IN^N$-graded
algebras long ago \cite{Rud}.

There might be alternatives to non-central extensions.
Figuerido and Ramos \cite{FR} suggested that associativity should
be abandoned, and presented some rather convincing arguments,
but nothing seems to have come out of this.

This paper is organized as follows. In \S 2 we define the $\ZN$-graded
Lie algebras of interest and recall their classical representations.
In \S 3 the notion of a ``core-central'' (but non-central) extension
is defined, which turns
out to be the natural generalization of central extensions to more than
one dimension. Explicit examples of core-central extensions are also
presented, which precisely generalize the Virasoro and affine
Kac-Moody algebras to
$N>1$. In \S 4 it is shown how to construct well-defined Fock modules
of certain core-central extensions of torus algebras. The energy of
these modules is bounded from below. It is
natural to ask whether these modules are irreducible. To this end,
we generalize in \S 5 Feigin's and Fuks' construction of singular
vectors to higher dimensions, modulo a technical assumption.
We thus have reducibility conditions for the extension of $\vect$.
The modules admitting the maximal number of simultaneous singular
vectors are especially important, but we have not managed to solve the
polynomial equations characterizing them.
Also, we have failed to obtain closed
formulas for the eigenvalues of the Cartan subalgebra in terms of
the Fock parameters, although it is clear that this can be done.
The final section contains some comments.

\sect{The classical representations}

Torus algebras act on fields, which can be expanded in a Fourier
basis
\be
\psi(x) = \sum_{n \in W} \psi(n) \ennx,
\ee
for some set of momenta $W$. This set is determined by the condition
that $\psi(x + 2 \pi \hat\jmath) = \eps \psi(x)$, where
$\hat\jmath$ is the unit vector in direction $j$ and
$\eps = +1$ for bosons and $\eps = -1$ for fermions.

\begin{definition}
\label{W}
Let $\one/2 = (1/2, 1/2, \ldots, 1/2) \in \ZN$,
$\La = \ZN \cup (\ZN + \one/2)$,
and $W = \ZN + v \subset \La$, where $v = 0$ for bosonic fields
and $v = \one/2$ for fermionic fields.
\end{definition}

$\La$ is the weight lattice of $\fcn$ and $\vect$,
$\ZN$ their root system and $W$ the set of momenta (weights) of
the class of modules corresponding to $v$.

Expanded in a Fourier basis, the semi-direct product between $\vect$
and the current algebra $\map$ takes the form
\bea
[L_\mu(m), L_\nu(n)] &=& n_\mu L_\nu(m+n) - m_\nu L_\mu(m+n),
\label{vect}
\lf
[L_\mu(m), J^b(n)] &=& n_\mu J^b(m+n),
\lf
[J^a(m), J^b(n)] &=& f^{abc} J^c(m+n),
\label{fcn}
\eea
where $m = (m^1, \ldots, m^N) \in \ZN$ and $f^{abc}$ are the
totally anti-symmetric structure constants of the finite-dimensional
Lie algebra $\oj$ equipped with a Killing metric $\delta^{ab}$.
$\fcn$ is a special case of $\map$ with $\oj$ abelian.

An important class of $\vect$ representations are
{\em tensor fields} (or densities). Let
\be
L_\mu(m) = \emx \Big( -i\d_\mu + w_\mu + m_\si T_\mu^\si \Big)
\label{tensor}
\ee
where $w_\mu$ is a constant vector defined modulo $\ZN$
and $\{T^\mu_\nu\}_{\mu,\nu=1}^N$ satisfies $gl(N)$, i.e.
\be
[T^\mu_\si, T^\nu_\tau] =
\delta^\nu_\si T^\mu_\tau -  \delta^\mu_\tau T^\nu_\si.
\label{glN}
\ee
It is straightforward to prove that (\ref{tensor}) satisfies
(\ref{vect}). Hence there is a $\vect$ representation for each
$gl(N)$ representation.
{}From a $gl(N)$ tensor with $p$ contravariant and
$q$ covariant indices and weight $\la$, the corresponding
tensor density is obtained.
Denote by $\TT^p_q(\la, w, v)$ the $\vect$ module with basis
$\{\psi^{\ind \si p}_{\ind \tau q}(n)\}_{n \in W}$,
with $W$ as in definition \ref{W}, and module action
\be\ba{l}
\disp{
[L_\mu(m), \psi^{\ind \si p}_{\ind \tau q}(n)] =
(n_\mu - w_\mu + (1-\la) m_\mu)
\psi^{\ind \si p}_{\ind \tau q}(m+n)
}
\lf
\disp{\qquad
+ \sum_{i=1}^p \delta^{\si_i}_\nu
\psi^{\indx \si p \nu}_{\ind \tau q}(m+n)
- \sum_{j=1}^q m_{\tau_j}
\psi^{\ind \si p}_{\indx \tau q \mu}(m+n).}
\ea\ee
We write $\psi \in \TT^p_q(\la, w, v)$ to indicate this formula.
In particular, the action on a scalar field is
\be
[L_\mu(m), \psi(n)] = (n_\mu - w_\mu + (1-\la) m_\mu) \psi(m+n),
\label{scalar}
\ee
or in position space,
\be
[L_\mu(m), \psi(x)] = - \emx (-i \d_\nu + w_\nu + \la m_\mu) \psi(x).
\ee
Define $\TT^p_q(\la) = \TT^p_q(\la, 0, 0)$.
The adjoint representation is $\TT_1^0(1)$.
The substitution
\be
\psi^{\ind \si p}_{\ind \tau q}(n) \tto
\psi^{\ind \si p}_{\ind \tau q}(n+v)
\ee
defines an isomorphism between the modules
$\TT^p_q(\la, w, v)$ and $\TT^p_q(\la, w-v, 0)$,
which thus may be identified.

There is a $\map$ representation for every $\oj$ irrep, which
extends to $\vect$ $\IX$ $\map$.
\be
[J^a(m), \psi(n)] = -M^a \psi(m+n),
\ee
where $[M^a, M^b] = f^{abc} M^c$ and representation indices
are suppressed.

The final classical representations are the {\em connections};
for $\vect$:
\bea
[L_\mu(m), \Ga^\si_{\tau\nu}(n)] &=&
(m_\mu + n_\mu) \Ga^\si_{\tau\nu}(m+n)
- m_\tau \Ga^\si_{\mu\nu}(m+n)
- m_\nu \Ga^\si_{\tau\mu}(m+n) \lf
&+& \delta^\si_\mu m_\rho \Ga^\rho_{\tau\nu}(m+n)
+ \delta^\si_\mu m_\tau m_\nu \delta(m+n)
\label{conn}
\eea
and for $\map$:
\be
[J^a(m), A^b_\nu(n)] = f^{abc} A^c_\nu(m+n) + m_\nu \delta(m+n).
\ee

\sect{Core-central extensions}

\begin{definition}
Let $\oj = \bigoplus_{m \in \ZN} \oj(m)$ be a $\ZN$-graded Lie
algebra.
An {\em extension} of $\oj$ is a $\ZN$-graded Lie algebra
$\toj$ containing the
brackets $[\oj(m), \oj(n)] \subset \oj(m+n) \oplus \hh(m,n)$,
where $\hh(m,n)$ in general has non-zero brackets with both
$\oj(s)$ and $\hh(s,t)$.
The extension is $\ZN$-graded by its {\em total momentum} $m+n$, and
its {\em core} is the part with zero total momentum, i.e.
$\bigoplus_{m \in \ZN} \hh(m,-m)$.
$\toj$ is a {\em core-central} extension of $\oj$ if its core is
central.
\end{definition}

The motivation for the name ``core'' is that it is the part of the
extension ``in the middle''.
Being central, the core can be added to the Cartan
subalgebra to characterize a representation. Of course, any central
extension is core-central, but there are also non-central extensions
with this property.
As is well known, torus algebras have no interesting central
extensions except in one dimension \cite{RSS,Lar1,Lar2}. Therefore,
the concept of a core-central extensions is a natural generalization
of central extensions to higher dimensions.
In this section some examples are given.

Every extension of $\vect$ has the form
\be
[L_\mu(m), L_\nu(n)] = n_\mu L_\nu(m+n) - m_\nu L_\mu(m+n)
+ R_{\mu\nu}(m,n),
\label{ext}
\ee
where $R_{\mu\nu}(m,n) = - R_{\nu\mu}(n,m)$.
Every extension of $\map$ has the form
\be
[J^a(m), J^b(n)] = f^{abc} J^c(m+n) + F^{ab}(m,n),
\ee
where $F^{ba}(n,m) = - F^{ab}(m,n)$.

The generic extension is local in the sense that the total momentum is
conserved, but it may
depend in an essential way on $m$ and $n$ separately.
However, there are examples of the form $f(m,n) S(m+n)$, where
$f(m,n)$ is an ordinary function.
The result of \cite{Lar1} is that the following abelian
(i.e. $[R_{\mu\nu}(m,n), R_{\si\tau}(s,t)] = 0$)
but non-central extensions are consistent with the Jacobi
identities.
\bea
R_{\mu\nu}(m,n) &=& m_\mu n_\nu (n_\si - m_\si) S^\si(m+n)
\label{S1}
\llf
R_{\mu\nu}(m,n) &=& m_\mu n_\nu m_\si n_\tau S^{\si\tau}(m+n)
\label{S2}
\llf
R_{\mu\nu}(m,n) &=& m_\pi m_\rho n_\si n_\tau
U_{\mu\nu}^{\pi\rho\si\tau}(m+n).
\eea
where the brackets with $L_\mu(m)$ are described by
$S^\si \in \TT^1_0(1)$, $S^{\si\tau} = -S^{\tau\si}
\in \TT^2_0(1)$ and $U_{\mu\nu}^{\pi\rho\si\tau} =
- U_{\nu\mu}^{\si\tau\pi\rho} \in \TT^4_2(1)$, respectively.

Denote by $C_p$ the $\TT^p_0(1)$ submodule consisting of
totally skew tensor densities, which may be identified
with the $p$-chains on the torus.
There is a $\vect$ homomorphism
\be
\begin{array}{ll}
\delta_p: & C_p \tto C_{p-1}
\lf
&(\delta_p S)^{\ind \nu {p-1}}(n)
= n_\si S^{\ind \nu {p-1} \si}(n).
\ea
\ee

The extension (\ref{S2}) may now be rewritten as
\be
m_\mu n_\nu m_\si n_\tau S^{\si\tau}(m+n)
= m_\mu n_\nu m_\si (\delta_2 S)^\si(m+n),
\ee
i.e. it is proportional to an exact one-chain. Similarly, we
may consistently demand that the one-chain in (\ref{S1}) is closed,
\be
(\delta_1 S)(n) = n_\si S^\si(n) = 0.
\label{closed}
\ee

Under the same assumptions, we have the following extension
of $\map$
\bea
F^{ab}(m,n) &=& \delta^{ab} (n_\si - m_\si) S^\si(m+n),
\label{mapext}
\llf
F^{ab}(m,n) &=& \delta^{ab} m_\si n_\tau S^{\si\tau}(m+n),
\eea

It is clear that (\ref{S1}, \ref{closed}, \ref{mapext})
together define non-central but core-central extensions. In one
dimension, (\ref{closed}) has only one solution: $S(n) = \delta(n)$,
and (\ref{S1}) and (\ref{mapext}) reduce to the Virasoro and affine
Kac-Moody algebras, respectively. Hence the notion of core-central
extensions naturally generalize central extensions to more than
one dimension.

We remark that some non-core-central extensions have appeared in
physics \cite{Mic,Fad}, e.g.
\be
F^{ab}(m,n) = d^{abc} m_\mu n_\nu F^{\mu\nu c}(m+n),
\qquad
d^{abc} = \tr \{J^a, J^b\} J^c.
\label{fadmic}
\ee

\sect{Fock modules}

To construct Fock modules of $\oj$ we must first define a
decomposition of $\oj$ into homogeneous components indexed by an
integer. For symmetrically $\IZ$-graded algebras this integer can
simply be identified with the degree. Torus algebras have a natural
$\ZN$-grading but there is no canonical bijection between $\ZN$
and $\IZ$. Therefore, a non-canonical choice must be made.

\begin{definition}
\label{energy}
Let $\La$ be as in definition \ref{W}.
An {\em energy function} is a function
$e: \La \tto \IR$, such that the following properties
hold for every $m, n \in \La$.
\begin{itemize}
\item[(i)]
$e(\cdot)$ is invertible.
\item[(ii)]
The number of momenta
$s \in \La$ with $e(m) \le e(s) < e(n)$ is finite.
\item[(iii)]
$e(0) = 0$ and $e(-m) = -e(m)$.
\item[(iv)]
If $e(m) > 0$, $e(n) > 0$, then $e(m+n) > 0$.
\end{itemize}
\end{definition}

However, the energy function is not additive: $e(m) + e(n) \ne
e(m+n)$ in general. Indeed, it is impossible to construct an
additive energy function when $N > 1$.
In one dimension the natural energy function is $e(n) = n$,
for every $n \in \La$.
An example for $N > 1$ can be constructed as follows. Split $\La =
\La_{(-)} \cup \{0\} \cup \La_{(+)}$ such that $m \in \La_{(+)}
\Rightarrow -m \in \La_{(-)}$. Assign to each momentum in $\La_{(+)}$
a unique positive integer $e(m)$, and extend this function to
$\La_{(-)}$ by $e(-m) = -e(m)$. Clearly, this can be done in many
ways.

\begin{definition}
\label{order}
Denote by $\lte$ the total order on $\La$ which is induced by
the energy function $e(\cdot)$: $m \lte n$ iff $e(m) < e(n)$.
\end{definition}

$\lte$ satisfies the axioms of an order relation (transitivity and
trichotomy), but it is not assumed to be additive:
$m \lte n$ does not imply $m+s \lte n+s$.

\begin{definition}
\label{voldef}
Denote by $\vol U$ the number of momenta in a subset $U \subset W$.
\end{definition}

Consider the vector space of scalar fermion fields on $\TN$,
with basis $\{ \psi(n) \}_{n \in W}$.
Because $\psi$ is fermionic, $W = \ZN + \one/2$ and $v = \one/2$;
this will henceforth always be the case.
The canonical anti-commutation relations (CAR) are
\bea
\{\psi(m), \psic(n)\} &=& \delta(m+n),
\lf
\{\psic(m), \psic(n)\} &=& \{\psi(m), \psi(n)\} = 0,
\label{CAR}
\eea
where $\psic$ is the conjugate of $\psi$.

For a given energy function there is a unique representation of the
CAR in a Fock space $\Fock(e)$ with vacuum vector
$\ket e$, such that
\be
\psi(n) \ket e = \psic(n) \ket e = 0,
\qquad \hbox{for $n \lte 0$.}
\label{vacuum}
\ee
The dual Fock space is defined by
\be
\bra e \psi(n) = \bra e \psic(n) = 0
\qquad \hbox{for $n \gte 0$.}
\ee

Each representation of a $\ZN$-graded algebra can be embedded
into the CAR algebra (the envelopping algebra of (\ref{CAR})).
For simplicity we discuss first $\oj = \fcn$.
Let
\be
J(m) = - \sum_{s \in W} \zeta \psic(m-s) \psi(s).
\label{embed}
\ee
It follows that
\be
[J(m), \psi(n)] = \zeta \psi(m+n),
\qquad
[J(m), \psic(n)] = -\zeta \psic(m+n).
\label{Jpsi}
\ee
We thus have a representation of $\fcn$ on the CAR algebra and a
representation of the latter on $\Fock(e)$. If $\fcn$ were
finite-dimensional it would inherit the $\Fock(e)$ module, but this
is not true here. The first problem is that the $J(0)$ eigenvalue
of the vacuum is infinite.
\be
J(0) \ket e = - \sum_{s \gte 0} \zeta \psic(-s) \psi(s) \ket e
= - \zeta \vol{\{ s: s \gte 0 \}} \; \ket e.
\ee
This problem is avoided by normal ordering. Redefine $J(0) \ket e
= j \ket e$ for some finite number $j$. In one dimension, this is
sufficient to render the module consistent, but not so when $N>1$.
An attempt to define a Fock module by
\be\ba{rcll}
J(m) \ket e &=& \disp{
- \sum_\su{s \gte 0}{ m-s \gte 0} \zeta \psic(m-s) \psi(s) \ket e,}
& m \gte 0,
\lf
&=& j \ket e, & m = 0,
\lf
&=& 0, & m \lte 0,
\ea
\label{infcent}
\ee
is inconsistent because there is an infinite central extension.
\be
[J(m), J(-m)] \ket e = - \zeta^2
\vol{\{s: s \gte 0 \and m-s \gte 0 \} } \; \ket e,
\ee
and there are infinitely many momenta $s$ that satisfy both
conditions, for every $m \gte 0$.

There is a remedy which seems natural, at least to this author.
Change the summation domain by replacing the condition $m-s \gte 0$
by $s \lte m$. Of course, the two conditions are equivalent if the
energy is linear, i.e. in one dimension, but otherwise the change
is quite dramatic. Namely, the energy function is defined so that
the modified summation domain, $0 \lte s \lte m$, is finite,
and hence no infinities can ever appear. This is the second
``renormalization'' needed to make the Fock construction consistent.
{}From a mathematical point of view, it is
no worse than normal ordering. In both cases the module is changed
by hand to make it well defined, at the expense of changing the
algebra as well.

There is also a physical argument why the modified algebra should have
the right ``classical limit''. A Fock vector is a superposition of
states of the form
\be
\psi(n_1) \ldots \psi(n_k) \psic(s_1) \ldots \psic(s_\ell) \ket e.
\label{class}
\ee
$\Fock(e)$ is an infinite-dimensional vector space, and hence
it is dominated by states where the number of quanta, i.e. $k$
and $\ell$, is very large. The action on such a state should depend
mostly on the classical commutatators (\ref{Jpsi}) and not so much
on the vacuum. Since (\ref{Jpsi}) has the right form, we expect that
the modified vacuum (and indeed, any reasonable vacuum)
gives rise to a good quantum version of $\fcn$. This hand-waving
argument does of course not prove that the resulting module is
physically relevant, but it is at least well defined.

\begin{definition}
\label{shiftfock}
Let $n_q \in W$ be the $q$:th momentum on $W$,
i.e. there are exactly $q-1$ momenta $n_j \subset W$
satisfying $0 \lte n_j \lte n_q$.
Define {\em shifted vacua} as follows.
\be
\ket{q; e} = \psi(n_q) \ldots \psi(n_2) \psi(n_1) \ket e,
\ee
if $q$ positive, and
\be
\ket{-q; e} = \psic(n_q) \ldots \psic(n_2) \psic(n_1) \ket e,
\ee
otherwise.
\end{definition}

To every energy function is associated a Hamiltonian $H$, satisfying
\be
[H, \psi(n)] = e(n) \psi(n), \qquad
[H, \psic(n)] = e(n) \psic(n), \qquad
H \ket e = 0.
\ee
Up to normal ordering,
\be
H = -\sum_{n\in W} e(n) \psic(-n) \psi(n) .
\label{Ham}
\ee
The corresponding energy is positive for every state in $\Fock(e)$.
The energies of the shifted vacua $\ket{q;e}$ and $\ket{-q;e}$ are
$\sum_{i=1}^q e(n_i) > 0$.

\begin{theorem}
\label{fockfcn}
The following expression defines, together with (\ref{Jpsi}), a
well-defined representation in $\Fock(e)$ of a certain core-central
extension of $\fcn$, denoted by $\tfcn$ .
\be\ba{rcll}
J(m) \ket e &=& \disp{
- \sum_{0 \lte s \lte m} \zeta \psic(m-s) \psi(s) \ket e,}
& m \gte 0,
\lf
&=& j \ket e, & m = 0,
\lf
&=& 0, & m \lte 0.
\ea
\label{fcndef}
\ee
This module has a decomposition
$\Fock(e) = \bigoplus_{q = -\infty}^\infty \Fock(q; e)$
into sectors with fixed fermion number.
$\Fock(q; e)$ is a $\tfcn$ module with lowest energy
(w. r. t. (\ref{Ham})), vacuum vector $\ket{q;e}$,
and $J(0)$ eigenvalue $j + q \zeta$.
\end{theorem}

\proof
By definition, the extension is $F(m,n) = [J(m), J(n)]$. Its
brackets are completely specified by (\ref{fcndef}):
\bea
[F(m,n), \psi(s)] &=& [F(m,n), \psic(s)] = 0,
\lf
F(m, n) \ket e &=& J(m) J(n) \ket e - J(n) J(m) \ket e.
\label{extdef}
\eea
In particular, if $m \gte 0$ and $n \lte 0$,
\bea
F(m,n) \ket e
&=& - \zeta^2 \sum_{A \setminus B} \psic(m+n-s) \psi(s) \ket e,
\lf
A &=& \{ s: 0 \lte s \lte m \and m+n-s \gte 0 \},
\lf
B &=& \{ s: 0 \lte s-n \lte m \and s \gte 0 \and m+n-s \gte 0 \}.
\eea
If the order is additive, the two sets are equal (they contain the
momenta satisfying $ 0 \lte s \lte m+n $) and the extension vanishes.
Since $F(m,n)$ does not otherwise vanish,
\be
[J(m), F(s,t)] = G(m,s,t),
\qquad
[F(m,n), F(s,t)] = H(m,n,s,t)
\ee
define two new extensions, whose brackets can be computed
analogously. The module is well defined because (\ref{fcndef}),
(\ref{extdef}) and (\ref{Jpsi}) only involve finite operations,
and hence only finite linear combinations appear.

The extension is core-central:
\be\ba{l}
\disp{
[J(m), J(-m)] \ket e =
J(-m) \sum_{0 \lte s \lte m} \zeta \psic(m-s) \psi(s) \ket e }
\lf \disp{
\qquad = \zeta^2 \sum_{0 \lte s \lte m} ( - \psic(-s) \psi(s)
+ \psic(m-s) \psi(s-m) ) \ket e }
\lf \disp{
\qquad - \zeta^2 ( \vol{\{s: 0 \lte s \lte n\}} -
\vol{\{s: 0 \lte s \lte m \and s-m \gte 0\}} ) \ket e. }
\ea
\label{jj}
\ee
If the energy function is additive, the second term vanishes because
the second condition is equivalent to $ s \gte m$, which clearly
has no overlap with the first condition.

$\Fock(e)$ can be decomposed because each generator $J(m)$,
$F(m, n)$, etc.,
preserves the fermion number (the difference between the number
of $\psi$'s and $\psic$'s).
The $J(0)$ eigenvalue is calculated thusly
\be
J(0) \ket {q;e} = [J(0), \prod_{i=1}^q \psi(n_i)] \ket e
+ \prod_{i=1}^q \psi(n_i) J(0) \ket e
= (q \zeta + j) \ket e.
\ee
\qed

We now turn to $\vect$.

\begin{theorem}
\label{fockvect}
The following expressions define a well-defined representation in
$\Fock(e)$ of a core-central extension of $\vect$, denoted by $\tvect$.
\be\ba{rcll}
L_\mu(m) \ket e &=& \disp{ - \sum_{0 \lte s \lte m}
(-\la m_\mu - w_\mu + s_\mu) \psic(m-s) \psi(s) \ket e,}
& m \gte 0,
\lf
&=& h_\mu \ket e, &m = 0,
\lf
&=& 0, & m \lte 0,
\ea\ee
and $\psi \in T^0_0(\la, w, \one/2)$,
$\psic \in T^0_0(1-\la, -w, \one/2)$.
This module has a decomposition
$\Fock(e) = \bigoplus_{q = -\infty}^\infty \Fock(q; e)$
into sectors with fixed fermion number.
$\Fock(q; e)$ is a $\tvect$ module with lowest energy
(w. r. t. (\ref{Ham})), vacuum vector $\ket{q;e}$,
and $L_\mu(0)$ eigenvalue $h_\mu - q w_\mu + \sum_{i=1}^q n_{i\mu}$.
\end{theorem}

\proof
Analogous to theorem \ref{fockfcn}.
\qed

Let us calculate the value the central core.
\be\ba{l}
[L_\mu(m), L_\nu(-m)] \ket e
\lf \disp{
\qquad= \sum_{0 \lte s \lte m}
(- \la m_\mu - w_\mu + s_\mu) ((1-\la)m_\nu + w_\nu - s_\nu)
\psic(-s) \psi(s) \ket e
} \lf \disp{
\qquad+ \sum_{0 \lte s \lte m}
(- \la m_\mu - w_\mu + s_\mu) ((\la-1) m_\nu - w_\nu + s_\nu)
\psic(m-s) \psi(s-m) \ket e
} \lf
\qquad= (- \alpha_{\mu\nu}(m) + \beta_{\mu\nu}(m) ) \ket e.
\ea\ee
where
\be\ba{rcl}
\alpha_{\mu\nu}(m) &=& \disp{
\sum_{0 \lte s \lte m}
(- \la m_\mu - w_\mu + s_\mu) ((\la-1) m_\nu - w_\nu + s_\nu), }
\lf
\beta_{\mu\nu}(m) &=& \disp{
\sum_\su{0 \lte s \lte m}{ m-s \gte 0}
(- \la m_\mu - w_\mu + s_\mu) ((\la-1) m_\nu - w_\nu + s_\nu). }
\label{tvpar}
\ea\ee
If the energy function can be chosen additively, the second term
vanishes because $s \gte m$, but otherwise not.

We have failed to express these functions in closed form in general,
but this may be easily achieved in the one-dimensional case.
\bea
\alpha_{11}(m) &=& 2 h m + {c \over {12}} (m^3 - m),
\lf
\beta_{11}(m) &=& 0,
\lf
c &=& -2 (6 \la^2 - 6 \la + 1) = 1 - 12 (\la - \half)^2,
\lf
2h &=& (w-q)^2 - (\la - \half)^2.
\label{virpar}
\eea
where $m = m_1$ is the only component of the vector $m$ and
$h = h_1$ is the $L(0)$ eigenvalue of $\ket{q;e}$.
The core-central extension is then central (the Virasoro algebra),
\be
[L(m), L(n)] = (n - m) L(m+n) - {c \over {12}} (m^3 - m) \delta(m+n).
\ee

Note that the value of $h_\mu$ is arbitrary; in one dimension,
it is fixed by demanding that the subalgebra generated by
$\{L(-1), L(0), L(1)\}$ has no extension. Similarly, the form
of the functions in (\ref{tvpar}) depends on $h_\mu$; it can
be fixed e.g. by demanding that there is no extension in
the $sl(N+1)$ subalgebra
generated by $\{ L_\nu(-\hat\nu), L_\nu(\hat\mu - \hat\nu),
\sum_{\si=1}^N L_\si(\hat\mu) \}_{\mu\nu=1}^N$, where $\hat\mu$
denotes a unit vector in the $\mu$ direction.

The modules constructed so far are manifestly consistent since
only finite polynomials in Fock space occur.
However, this may be too restrictive, because the physical
condition is only that all matrix elements are finite. The
following modules are well defined in this weaker sense.

\begin{theorem}
\label{fd}
Consider the $\tfcn$ and $\tvect$ modules defined in theorems
\ref{fockfcn} and \ref{fockvect}.
Replace, for $m \gte 0$, the action of $J(m)$ and $L_\mu(m)$
on $\ket e$ by
\be\ba{rcl}
J(m) \ket e &=& \disp{
- \sum_\su{s \gte 0}{m-s \gte 0} \ffd{e(s) - e(m)}{\beta} \;
\zeta \psic(m-s) \psi(s) \ket e, }
\lf
L_\mu(m) \ket e &=& \disp{
- \sum_\su{s \gte 0}{m-s \gte 0} \ffd{e(s) - e(m)}{\beta} \;
(-\la m_\mu - w_\mu + s_\mu) \psic(m-s) \psi(s) \ket e, }
\ea\ee
where $\ffd{\eps}{\beta} = 1 / (1 + \exp(\beta \eps))$
is the Fermi-Dirac distribution function and $\beta$ is a positive
parameter.
These expressions define representations in $\Fock(e)$ of
certain core-central
extensions of $\fcn$ and $\vect$, denoted by $\tfcnb$ and $\tvectb$.
Although infinitely many terms are created out of the vacuum,
every matrix element is finite.
The representations decompose into sectors with fixed fermion number
and vacua $\ket{q; e}$.
\end{theorem}

\proof
It is clear that the extensions are core-central and that the fermion
number is conserved. Hence the only thing left to prove
is that all matrix elements are finite. Consider
a typical matrix element
\be\ba{l} \disp{
\bra e J(m_1) \ldots J(m_k) \ket e
= (-\zeta)^k \sum_{s_1 \ldots s_k}
\ffd{e(s_1) - e(m_1)}\beta \ldots \times}
\lf \qquad
\ffd{e(s_k) - e(m_k)}\beta
\bra e \psic(m_1 - s_1) \psi(s_1) \ldots \ket e.
\ea\ee
The only possible cause of divergence is that the sum runs
over infinitely many momenta. However, the number of momenta with
total energy $\eps = e(s_1) + \ldots + e(s_k)$ grows only polynomially
in $\eps$, whereas $\ffd{\eps}\beta \approx \exp(-\beta \eps)$
falls off exponentially fast. The sum thus converges and the matrix
element is finite.
\qed

If $\eps$ is kept fixed, the Fermi-Dirac distribution has the limits
$\ffd\eps\beta \to \theta(-\eps)$ when $\beta \to \infty$ and
$1/2$ when $\beta \to 0$. Hence we have, formally,
$\tfcn = \lim_{\beta \to \infty} \tfcnb$, whereas
$\lim_{\beta \to 0} \tfcnb$ is the inconsistent module (\ref{infcent})
with infinite central extension, up to normalization.
We may thus view the family of Fock modules in Theorem \ref{fd} as
a regularized version of the na\"\i ve Fock construction. All
observables will depend analytically on $\beta$, except possibly in the
limit $\beta \to 0$. However, if this limit exists, it can be used
to define observables for $\beta = 0$.
Analogous results hold for $\vect$.

$\beta$ plays the role of an inverse temperature and $e(m)$ that of
a chemical potential. It is not clear to us if this has a physical
interpretation, because we introduced the Fermi-Dirac
distribution solely as a mathematical trick to avoid divergencies.

Analogous Fock modules exist for the current
algebra $\map$. In this case the root system is
$\ZN \times \Phi_\oj$, where $\Phi_\oj$ is the root system of
$\oj$.
The Fock construction gives rise to a certain core-central extension,
which in one dimension is central: the
affine Kac-Moody algebra.

\sect{Reducibility conditions for \vect}

Feigin's and Fuks' celebrated construction of singular vectors in
the Virasoro
algebra consists of two steps: find invariants in the CAR algebra
and apply these invariants to the vacuum. Their construction
is generalized to $\tvectb$ in this section.

\begin{lemma}
\label{invariant}
If $\Psi \in \TT^0_0(1, w, v)$, $\Psi(w)$ is an $\vect$ invariant.
\end{lemma}

\proof
$[L_\mu(m), \Psi(n)] = (n_\mu - w_\mu) \Psi(m+n)$.
\qed

In view of this lemma, we must find a way to construct scalar
densities with weight $\la = 1$ in order to construct invariants.
To this end, we use that pointwise multiplication and the
exterior derivative are $\vect$ homomorphisms, the latter depending
on the connection (\ref{conn}).
\be\ba{ll}
*: &\TT^{p_1}_{q_1}(\la_1, w_1, v_1)
\ot \TT^{p_2}_{q_2}(\la_2, w_2, v_2)
\tto \TT^{p_1+p_2}_{q_1+q_2}(\la_1+\la_2, w_1+w_2, v_1+v_2)
\lf
\del: &\TT^p_q(\la, w, v) \tto \TT^p_{q+1}(\la, w, v),
\ea\ee
E.g., the covariant derivative of a scalar field is
\be
(\del\psi)_\mu(x)
= (\d_\nu + iw_\nu + \la \Ga_{\si\nu}^\si(x)) \psi(x).
\ee

\begin{definition}
A {\em composite field} is the field obtained by multiplying various
covariant derivatives of $\psi(x)$ at the same point $x$.
The $k$:th {\em shell} of a composite field is the factor containing
the $k$:th derivatives of $\psi(x)$.
The {\em occupation number} of the $k$:th shell is the number
$p_k$ of factors of $\psi$ in this shell.
\end{definition}

The usefulness of this definition is that every composite field is
a tensor field.
The general expression for composite fields is
rather cumbersome in $N$ dimensions, but when $N=1$ it reads
\be
\Phi^{(p_0, p_1, p_2, \ldots)}(x)
= \psi(x)^{p_0} (\del \psi(x))^{p_1} (\del^2 \psi(x))^{p_2} \ldots.
\ee
The $k$:th shell is thus the factor $(\del^k \psi(x))^{p_k}$,
$p_k$ is the occupation number of this shell,
and only finitely many $p_k$ are non-zero.
Since $\psi$ is fermionic $p_k \le 1$;
if $p_k = 1$ we say that the $k$:th shell is filled, if $p_k = 0$
it is empty.

A composite field depends in general on the connection, but when the
first $p$ shells are filled and the rest empty
($p_k = 1, k < p$, and $p_k = 0, k \ge p$), this is not the case.
Denote this special composite field by $\Psi^{(p)}(x)$. E.g.,
\bea
\Psi^{(2)}(x) &=& \Phi^{(1, 1, 0, 0, \ldots)}(x)
= \psi(x) (\d + iw + \la \Ga(x)) \psi(x)
\lf
&=& \psi(x) \d \psi(x) + (iw + \la \Ga(x)) \psi(x)^2,
\eea
and the second term vanishes because $\psi(x)^2 = 0$. $\Psi^{(p)}(x)$
is simply the $p$-fermion Vandermonde determinant.

In higher dimensions the occupation numbers may be larger than one,
because the covariant derivative has several components. We say that
a shell is filled if its occupation number is maximal; this number
depends on the shell. If only the $p+1$ shells are  non-empty and
the first $p$ shells are filled, the composite field is
independent of the connection.
In two dimensions, the composite field with $p=3$ filled shells is
\be
\ba{l}
\Psi^{(3)}_{\nu_1 \nu_2 \si_{11} \tau_{11} \si_{12}
\tau_{12} \si_{22} \tau_{22}}(x)
\lf
\qquad= \psi(x) \, (\del\psi)_{\nu_1}(x) \, (\del\psi)_{\nu_2}(x)
\, (\del\del\psi)_{\si_{11} \tau_{11}}(x)
\, (\del\del\psi)_{\si_{12} \tau_{12}}(x)
\, (\del\del\psi)_{\si_{22} \tau_{22}}(x)
\lf
\qquad = \psi(x) \, \d_{\nu_1} \psi(x) \, \d_{\nu_2} \psi(x)
\, \d_{\si_{11}} \d_{\tau_{11}} \psi(x)
\, \d_{\si_{12}} \d_{\tau_{12}} \psi(x)
\, \d_{\si_{22}} \d_{\tau_{22}} \psi(x).
\ea
\label{3shell}
\ee
To arrive at the last expression, we used that all references to the
connection vanishes due to anti-symmetry, and thus all covariant
derivatives can be replaced by ordinary ones.

\begin{lemma}
\label{tensormap}
There is a map
\be
\La^{A_N(p)} \TT^0_0(\la, w, \one/2)
\tto \TT^0_{B_N(p)}(A_N(p) \la, A_N(p) w, A_N(p) \one/2),
\ee
where
\be
A_N(p) = {{N-1+p} \choose N},
\qquad
B_N(p) = N {{N-1+p} \choose {N+1}}.
\ee
\end{lemma}

\proof
Consider the composite field $\Psi^{(p)}(x)$ with $p$ filled shells.
The occupation number of the $k$:th shell ($k < p$) is
equal to the number of symmetric combinations
of $k$ indices which can take $N$ different values, i.e.
\be
p_k = {{N-1+k} \choose k} = {{N-1+k} \choose {N-1}}.
\label{p_k}
\ee
Each $\psi$ in the $k$:th shell
contributes with $k$ lower indices, wherefore
\be
\Psi^{(p)} \in \TT^0_{B_N(p)}(A_N(p) \la, A_N(p) w, A_N(p) \one/2),
\ee
where
\bea
A_N(p) &=& \sum _{k=0}^{p-1} p_k
= \sum _{k=0}^{p-1} {{N-1+k} \choose {N-1}}
= {{N-1+p} \choose N},
\lf
B_N(p) &=& \sum _{k=0}^{p-1} k p_k
= \sum _{k=0}^{p-1} k {{N-1+k} \choose {N-1}}
= N \sum _{j=0}^{p-2} {{N+j} \choose N}
\lf
&=& N {{N-1+p} \choose {N+1}}.
\eea
The domain of the map is determined by the fact that $A_N(p)$ is
the total number of $\psi$'s in the $p$-shell composite field.
\qed

The range of this map is actually a submodule, characterized by
certain symmetries. The composite field (\ref{3shell})
is skew in $\nu_1$ and $\nu_2$, symmetric in $\si_{ij}$ and
$\tau_{ij}$, and skew under interchange of any pairs
$ \si_{ij} \tau_{ij} \leftrightarrow \si_{kl} \tau_{kl}$.

{}From lemma \ref{tensormap} we obtain a composite field, which has
lower indices with certain symmetries. It can
transformed into a scalar density by multiplication of some field with
upper indices. There is a canonical choice: the permutation symbol
may be regarded as a certain tensor density.

\begin{lemma}
There is a totally skew constant field
$\eps^{\nu_1 \ldots \nu_N}(x) \in \TT^N_0(1)$,
defined by $\eps^{1 2 \ldots N}(x) \equiv 1$.
\end{lemma}

\proof
By a direct computation, using the transformation law of
$\TT^N_0(1)$, skewness, and constancy, it is found that
$ [L_\mu(m), \eps^{\nu_1 \ldots \nu_N}(x)] = 0$. Hence the
assumptions are consistent.
\qed

Dually, we may regard the permutation symbol as a constant field in
the
module $\TT^0_N(-1)$. When $N=1$, $\eps(x) = 1$ is the invariant
of the module $\TT^0_0(1)$ (lemma \ref{invariant}).

The desired scalar field is now formed by contracting the composite
field in lemma \ref{tensormap}
with the permutation symbol in a way which respects the symmetries.
{}From (\ref{3shell}), we obtain the following scalar field.
\bea
&&\eps^{\nu_1 \nu_2}
\, \eps^{\si_{11} \si_{12}}
\, \eps^{\tau_{11} \si_{22}}
\, \eps^{\tau_{12} \tau_{22}}
\, \psi \, \d_{\nu_1} \psi \, \d_{\nu_2} \psi
\, \d_{\si_{11}} \d_{\tau_{11}} \psi
\, \d_{\si_{12}} \d_{\tau_{12}} \psi
\, \d_{\si_{22}} \d_{\tau_{22}} \psi
\lf
&\propto& \psi \, \d_1 \psi \, \d_2 \psi \, \d_1 \d_1 \psi
\, \d_1 \d_2 \psi \, \d_2 \d_2 \psi.
\eea
This procedure is well defined for
arbitrary composite fields with full shells.

\begin{lemma}
\label{scalarmap}
There is a map
\be
\La^{A_N(p)} \TT^0_0(\la, w, \one/2) \longrightarrow
\TT^0_0(A_N(p) \la + {1 \over N} B_N(p), A_N(p) w, A_N(p) \one/2).
\ee
\end{lemma}

\proof
There are $B_N(p)$ lower indices to contract.
Since each $\eps^{\nu_1 \ldots \nu_N}$
has $N$ upper indices, a total of $B_N(p)/N$ permutation symbols is
needed, each  contributing unity to the parameter $\la$ and nothing
to $w$.
\qed

\begin{proposition}
\label{reducible}
There is an invariant in the module
$\La^{A_N(p)} \TT^0_0(\la, w, \one/2)$ provided that
$A_N(p)$ is even and
\bea
A_N(p) \la + {1 \over N} B_N(p) &=& 1,
\lf
A_N(p) w_\mu &=& n_\mu,
\label{red}
\eea
for some $n \in \ZN$.
\end{proposition}

\proof
Combine lemmas \ref{invariant} and \ref{scalarmap} and note that
$A_N(p) = 0 \bmod 2$.
\qed

The reducibility condition (\ref{red}) can be cast into
different equivalent forms.
\bea
\la - {1 \over {N+1}} &=& {1 \over {A_N(p)}} - {p \over {N+1}},
\lf
w_\mu &=& {{n_\mu} \over {A_N(p)}},
\eea
because $B_N(p) = N (p-1) A_N(p) / (N+1)$.
Every solution to the second equation is parallel to $w_\mu$.
If we introduce $\kappa = (N+1) \la - 1$, the first equation takes
the form
\be
(p + \kappa) {{N+p-1} \choose N} = N+1,
\ee
i.e.
\be
p (p+1) \ldots (p+N-1) (p + \kappa) = (N+1)!.
\label{pcond}
\ee
This is a polynomial equation of degree $N+1$, which generically has
$N+1$ complex solutions $p_i$. The maximal number of invariants in
this module is thus $N+1$,
which is obtained if all $p_i$ are real and different and
satisfy the simultaneous Diophantine equations
\be
w_\mu = {{n_{i\mu}} \over {A_N(p_i)}},
\qquad
n_{i\mu} \in \ZN,
\qquad i = 1, \ldots, N+1.
\label{wcond}
\ee

For concreteness we explicitly list the invariant conditions
(\ref{pcond}), (\ref{wcond}) for $N = 1, 2, 3$.
For $N=1$,
\be
p\la + {{p(p-1)} \over 2} = 1,
\qquad
pw = n,
\label{1red}
\ee
with the solutions
\be
p_{1,2} = -(\la-\half) \pm \sqrt{(\la - \half)^2 + 2}.
\ee
There are two different invariants provided that there are two
integers
$n_1$ and $n_2$ such that $p_i = n_i x$ ($x = w^{-1}$), i.e.
\be
n_1 n_2 x^2 = -2,
\qquad\qquad
(n_1 + n_2) x  = -\kappa.
\ee
This equation has a solution if
\be
n_1 n_2 < 0,
\qquad
x = \sqrt{-{ 2 \over {n_1 n_2}}},
\qquad
\kappa^2 = -{{2 {(n_1 + n_2)}^2} \over {n_1 n_2}}.
\label{1max}
\ee

$N=2$:
\bea
1 + \kappa &=& -(p_1 + p_2 + p_3),
\lf
\kappa &=& p_1 p_2 + p_1 p_3 + p_2 p_3,
\lf
6 &=& p_1 p_2 p_3,
\eea
where the $p_i$ are related by
\be
{{p_i (p_i+1)} \over 2} = n_i x,
\qquad n_i \in \IZ.
\ee

$N=3$:
\bea
3 + \kappa &=& - \sum_{i=1}^4 p_i,
\lf
2 + 3\kappa &=& \sum_{1\le i < j \le4} p_i p_j
\lf
2 \kappa &=& - \sum_{1\le i < j < k \le4} p_i p_j p_k
\lf
24  &=& - p_1 p_2 p_3 p_4,
\eea
where the $p_i$ are related by
\be
{{p_i (p_i+1) (p_i+2)} \over 6} = n_i x,
\qquad n_i \in \IZ.
\ee

We have not been able to find any solution to these simultaneous
equations for $N > 1$.

The next step in the Feigin-Fuks procedure consists of applying an
invariant to the Fock vacuum. The resulting Fock vector is singular.

\begin{definition}
A {\em singular vector} $\sing \in \Fock(e)$
is annihilated by every generator with negative total momentum.
It thus satisfies the same conditions as the vacuum, i.e.
$L_\mu(m) \sing = 0$ for $m \lte 0$, etc.
\end{definition}

\begin{conjecture}
\label{sing}
Let $\Psi^{(p)}(n) \in \TT^0_0(1, n, A_N(p) \one/2)$, where
$n \in \ZN$ and $A_N(p)$ is even,
be the $\vect$ invariant constructed in
proposition \ref{reducible}. Then
\be
\sing = (\Psi^{(p)}(n))^j \ket{e}
\ee
is a singular vector in the reducible $\tvectb$ module
$\Fock(j A_N(p); e)$.
\end{conjecture}

\remark
Because $A_N(p)$ is even we can, at least formally, continue
to half-integers and set $j = s/2$, $s \in \IN$.

\proof
The vector is singular:
\be
L_\mu(m) \sing
= j [L_\mu(m), \Psi^{(p)}(n)] (\Psi^{(p)}(n))^{j-1} \ket{e}
+ (\Psi^{(p)}(n))^j L_\mu(m) \ket{e}.
\ee
The first term vanishes because $\Psi^{(p)}(n)$ is invariant.
Hence $\sing$ is a singular vector because $\ket e$ is so.
It is a vector in
$\Fock(j A_N(p); e)$, which thus is reducible.
\qed

However, one difficulty remains, which is why we consider this
statement to be a conjecture. The singular vector
is a sum of infinitely many terms. Indeed, it can be written in the
form
\be\ba{l} \disp{
\sing = \bigg[ \sum_\su{s_i \gte 0}{\Sigma_i s_i = n}
\Delta(s_1, \ldots, s_{A_N(p)}) \, \psi(s_1) \ldots \psi(s_{A_N(p)})
\bigg]^j \ket e, }
\ea\ee
for some function $\Delta$.
Despite the constraint $\sum_i s_i = n$, the sum runs over
infinitely many momenta $s_i > 0$, except in one dimension.

Nevertheless, we believe this difficulty to be of technical
nature only. For example, a new norm in Fock space could be defined,
unrelated to the inner product, such that $\sing$ is in the closure
of $\Fock(e)$ relative to this norm.
This can be done by damping the contribution from monomials with
total momentum $m$ by $\exp(- |e(m)|)$.
The new norm can be relevant if we consider the algebra of
smooth vector fields,
$L_\xi = \sum_m \xi_\mu(-m) L_\mu(m)$, where
$\xi_\mu(m)$ falls off exponentially with $|e(m)|$.

Assume that it is possible to make sense of conjecture \ref{sing}.
A large number of non-trivial modules can be constructed as follows.
Introduce the equivalence relation $\sing \sim 0$.
Then the factor module $\Fock_1 = \Fock(j A_N(p); e) / \sim$
is well-defined and non-trivial.
There might exist another singular vector in $\Fock_1$; if so,
$\Fock_1$ can be further reduced to $\Fock_2$ by equalling
this vector to zero. This procedure can be carried on until all
singular vectors are exhausted.

Unfortunately, we have not yet been able to arrive at explicit formulas
for the main parameters characterizing the maximally reducible
modules. One problem is that we lack solutions to
(\ref{pcond}) and (\ref{wcond}). However, this is not so serious,
because the modules which admit some but not all of the invariants
are still exceptionally small.
Another difficulty is that the relation between $\la$ and $w$ and
the functions (\ref{tvpar}) is complicated and depends on the
energy function. This problem can be solved as follows. Proclaim that
the extension should vanish in a certain finite-dimensional subalgebra
$\oj$. Choose an order such that the generators of $\oj$ are
first, which essentially amounts to a suitable choice of energy function.
The restriction of $\tvect$ yields a proper representation of $\oj$ in
Fock space, and the parameters $\la$ and $w$ can be related to the
eigenvalues of the Cartan subalgebra. In particular, if we reinterpret
$\vect$ as the derivation algebra of Laurent polynomials, $\oj$ can
be chosen as the conformal algebra. Hence a discrete spectrum of
preferred critical exponents is obtained.
We will address this issue in a future publication.

In one dimension we can proceed further. The reducibility
conditions in terms of $\la$ and $w$ were calculated in (\ref{1red}),
and the relation between these parameters and $h$ and $c$ follows
from (\ref{virpar}), where $q = jp = sp/2$.
\bea
c &=& 1 - 12 {\Big({1\over p} - {p \over 2} \Big)}^2,
\lf
2h &=& {\Big( {n \over p} - {{sp} \over 2} \Big)}^2
- {\Big( {1\over p} - {p \over 2} \Big)}^2.
\eea
By eliminating $p$ we obtain for
the special values of $\la$ given by (\ref{1max}),
\bea
c &=& 1 - {{6{(k - l)}^2} \over {kl}}
\lf
h &=& {{{(nk-sl)}^2 - {(k-l)}^2} \over {4kl}}
\label{discrete}
\eea
where $k = n_1$, $l = - n_2$, $k, l, n, s \in \IN$.
This is the discrete series
of irreducible Virasoro representations which have many applications
in physics \cite{ISZ}.

\sect{Conclusion}

In this paper manifestly consistent lowest-energy representations
of core-central extensions of torus algebras have been constructed.
The main new ingredient is the ``renormalization'',
which renders the modules finite.
Previously the problem was that Fock modules could not be constructed.
Now the problem is the opposite; there are too many of them.
Different energy functions and renormalization prescriptions give
rise to non-equivalent Fock modules.
On the other
hand, there are only a few main parameters such as $\la$ and $w$,
and the reducibility conditions only depend on these. There is thus
a kind of universality; a few universal quantities which are the same
throughout every universality class. The main problem is clearly to
calculate universal numbers in irreducible representations.
We believe that such numbers appear in nature.

The present work can be extended in various ways.
The form of the core-central extensions, or
at least their cores, should be explicitly calculated for some
specific choice of energy function. Since $\IC \subset \IR^2$,
we expect that the
Virasoro algebra is a subalgebra of $\tvectbtwo$ for some $\beta$,
but this must be verified.
Only Fock modules
based on scalar densities have been considered, but the generalization
to arbitrary tensor fields is straightforward. The construction
of invariants and singular vectors also goes through with obvious
modifications. However, one new feature arises. Consider the
bosonic Fock module constructed from a metric $g^{\mu\nu} \in
\TT^2_0(0)$, and let $G_{\mu\nu}(n)$ be the Fourier components of
the corresponding Einstein tensor. The condition
$G_{\mu\nu}(n) \ket e = 0$ for every $n$ can consistently be imposed,
modulo technical problems analogous to conjecture \ref{sing}.
The factor module has a lowest energy and solves the
Einstein equation in empty space, wherefore it might be relevant for
quantum gravity.

\end{document}